\documentclass[showpacs,groupedaddress,assymb,amsmath]{revtex4}
\usepackage{graphicx}
\usepackage{mathrsfs}
\usepackage{amssymb}
\usepackage{amsmath}

\begin{document}

\title{Preparation of edge states by shaking boundaries}
\author{Z. C. Shi$^{1,2}$, S. C. Hou$^{3}$,  L. C. Wang$^{4}$,
and X. X. Yi$^{2}$\footnote{Corresponding  address:
yixx@nenu.edu.cn}}
\affiliation{$^1$ Department of Physics, Fuzhou University, Fuzhou 350002, China\\
$^2$ Center for Quantum Sciences and School of Physics, Northeast Normal University, Changchun 130024, China \\
$^3$ Institute of Fluid Physics, China Academy of Engineering Physics, Mianyang, Sichuan, China\\
$^4$ School of Physics and Optoelectronic Technology Dalian University of Technology, Dalian 116024 China}

\date{\today}

\begin{abstract}
Preparing topological states of quantum matter, such as edge states,
is one of the most important directions in condensed matter physics.
In this work, we present a proposal to prepare edge states in
Aubry-Andr$\acute{\textrm{e}}$-Harper (AAH) model with open
boundaries, which takes advantage of Lyapunov control to design
operations. We show that edge states can be obtained with almost arbitrary
initial states. A numerical optimalization for the
control is performed and the dependence of control process on the system size
is discussed. The merit of this proposal  is that the
shaking exerts only on the boundaries of the model. As a by-product,
a topological entangled state is achieved by elaborately designing
the shaking scheme.
\end{abstract}
 \maketitle

\section{introduction}
Topological insulators(TIs) \cite{hasan10,qi11} are new states of
quantum matter with profound physical features that have their
origin in topology. They have a bulk insulating gap but are distinct
from ordinary insulators  at stable gapless edge states. The quantum
Hall system\cite{thouless82}, which is an insulator without any kind
of spontaneous symmetry breaking, is the first example of
topological insulators.

It is believed that there only exists topological trivial phase in
1D systems due to  the lack of symmetries. Recently, it has been
shown that the 1D quasiperiodic system \cite{YEK1109} shares the
same topological non-trivial phase emerged in the 2D integer quantum
Hall effect \cite{DRH-PRB14}. Indeed, the edge state that
characterizes topological property in this system has been observed
\cite{YEK1109}. Most recently, topologically protected edge
state has been demonstrated in the 1D commensurate off-diagonal
Aubry-Andr$\acute{\textrm{e}}$-Harper (AAH) model \cite{SA-AIPS3},
interpreted by the $\textrm{Z}_2$ topological index of the Kitaev
model \cite{AYK-PU44}. It is worth noticing that the topology
properties have been studied in  optical lattices extensively \cite{
hormozi12,degottardi13,satija13, cooper12, price12, kjall12,
atala13,zhu13, grusdt13,
deng13,NG-PRL105,TDS-PRA82,BB-PRL107,JK-PRL111,ganeshan13}, and the edge states
have been observed in the photonic system
\cite{TK1105,MH1302}.

State steering is an important task in quantum control. Many
protocols \cite{DD2007} have been proposed for this purpose,
including the optimal control technique, adiabatic control, the
technique of stimulated Raman scattering involving adiabatic passage
(STIRAP), and  open loop controls based on Lyapunov functions. Among
them, the Lyapunov based control has its own advantages  due to the
effectiveness designing for control fields. To apply  the
Lyapunov control, a function called Lyapunov function has to be
specified first.  Then the control fields can be designed by guaranteeing
the decrease of Lyapunov function, and the system would evolve to the target state
asymptotically. In particular, this control scheme has been studied by several
researchers
\cite{KB-SCL56,SK-A44,XXY-PRA80,XTW-PRA80,JMC-NJP11,WW-PRA82,XTW-IEEE55,SCH-PRA86}
and has been applied  to diverse fields  in physics.

In this paper, we steer an arbitrary given initial state
into the edge states in a 1D optical lattice by  only shaking
the on-site energy of the boundaries, which can be established via the Lyapunov
control. By choosing different Lyapunov functions, we show distinct
convergence behaviors for the system. In addition, we also explore a
feasible way to realize boundary-boundary entangled states \cite{LCV-PRA76,SID-JETP85,LB-PRA82}.

The paper is organized as follows. In Sec. {\rm II}, we  present a
general formalism  for the Lyapunov control. In Sec. {\rm III},  we
first apply the general method to steer the system into edge state
by shaking  only the energy of boundary sites, and study the
dynamical behaviors with different Lyapunov
functions. Then we show how to optimize the Lyapunov function in
terms of fidelity and control time. An exploration on the effect of
errors on the fidelity of final state is given in Sec. {\rm IV}. By
use of the elaborate designing control fields, we study the
behavior of boundary-boundary entangled states in Sec. {\rm
V}. Finally, we conclude in Sec. {\rm VI}.

\section{Lyapunov control}

Although the  problem of quantum control might be formulated in
different ways, the final purpose  is mostly to steer a quantum
system from an (arbitrary)  initial state to a target state by
control fields. We start with the dynamics of a  quantum
system governed by Schr\"odinger equation
($\hbar=1$)
\begin{eqnarray}\label{15}
|\dot{\psi}(t)\rangle&=&-i[H_0+\sum_{k} f_{k}(t)H_{k}]|\psi(t)\rangle,
\end{eqnarray}
where $H_0$ and $H_k$ denote the system Hamiltonian and control Hamiltonian, respectively.
$f_k(t)$ represent the control fields. As mentioned in the
introduction, the design of the control fields can be different from
proposal to proposal. Here, we use the Lyapunov control
technique to design the control fields. The essence of the Lyapunov
control is to choose a Lyapunov function, which is required to be
positive and reaches its minimum when the system arrives
at target state.  Obviously, the following form of Lyapunov
function
\begin{eqnarray}\label{2}
V_{1}=1-|\langle \psi_T|\psi(t)\rangle|^2,
\end{eqnarray}
meets the requirement.  Here $|\psi_T\rangle$ denotes the  target
state which is conventionally an eigenstate of system Hamiltonian.
The time derivative of $V_1$ yields
\begin{eqnarray}\label{3}
\dot{V}_1
   =-2\sum_{k}|\langle  \psi(t)|\psi_T\rangle|f_k(t)\textrm{\textbf{Im}}
   [e^{i\arg\langle\psi(t)|\psi_T\rangle}
\langle \psi_T|H_k|\psi(t)\rangle],
\end{eqnarray}
where $\textrm{\textbf{Im}}[\cdot]$ stands for the imaginary part of
$[\cdot]$ and $\textrm{arg}\langle\psi(t)|\psi_T\rangle$ is the angle
between states $|\psi(t)\rangle$ and $|\psi_T\rangle$. Thus, the
condition $\dot{V}_1\leq0$ can be satisfied naturally if we choose
the control fields
$f_k(t)=A_{1k}\textrm{\textbf{Im}}[e^{i\arg\langle\psi(t)|\psi_T\rangle}
\langle \psi_T|H_k|\psi(t)\rangle]$ with $A_{1k}>0$.  Especially, $A_{1k}$  can be
used to adjust the amplitude of control fields and the
control time.

As the Lyapunov function is not unique, it can be
constructed differently even though the control problem is same.
For example, we may  define  another Lyapunov function by
\begin{eqnarray}\label{4}
V_{2}=\langle\psi(t)|P|\psi(t)\rangle,
\end{eqnarray}
where the operator $P$ is hermitian and time independent.
Additionally, $P$ is also assumed to be positive semidefinite
operator acting on the Hilbert space spanned by the eigenvectors of
the system Hamiltonian $H_0$. With this definition, the time derivative of
$V_2$ becomes
\begin{eqnarray}\label{5}
\dot{V}_2&=&\langle \psi(t)|i[H_0,P]|\psi(t)\rangle+\sum_{k}f_k(t)\langle
         \psi(t)|i[H_k,P]|\psi(t)\rangle.
\end{eqnarray}
The Lyapunov control requires that the operator  $P$ should  be
constructed properly to guarantee \cite{XXY-PRA80}
\begin{eqnarray}\label{15}
[H_0,P]=0.
\end{eqnarray}
Therefore a straightforward way to construct the operator $P$ is
\begin{eqnarray}\label{Lya2}
P=p_f|\lambda_f\rangle\langle\lambda_f|+\sum_{i\neq
f}p_i|\lambda_i\rangle\langle\lambda_i|,
\end{eqnarray}
where $|\lambda_i\rangle$ ($i=1,...,N$) are the eigenstates of system
Hamiltonian with  corresponding eigenvalues $\lambda_i$, and
$|\lambda_f\rangle$ is the target state. Of course, the values of
$p_f$ and $p_i$ ($i=1,...,N, i\neq f$) can be chosen  arbitrarily
except for the necessary condition $p_f<p_i$\cite{XXY-PRA80}.
Clearly, if we select the control fields $f_k(t)=-A_{2k}\langle
\psi(t)|i[H_k,P]|\psi(t)\rangle$ with $A_{2k}>0$, the condition
$\dot{V}_2\leq0$ is satisfied naturally. Discussions on the Lyapunov
function are in order. The Lyapunov function given in equation
(\ref{4}) with equation (\ref{Lya2}) is quite general for unitary
systems. In fact, the Lyapunov function $V_2$ covers the Lyapunov
function $V_1$ \cite{note2}.  This suggests that the optimalization
over the Lyapunov function reduces to searching a set of $\{p_j\}$
that maximizes the fidelity of final state or minimizes the control
time, etc., depending on the optimal function. We will carry out
this optimalization latter.

According to the Lyapunov control theory, the Lyapunov  function
$V_k$ will converge to its minimum while the state of system
converges to a LaSalle's invariant set given  by $\mathcal
{E}_k=\{\dot{V_k}=0\}$ ($k=1,2$). When the dimension of LaSalle's invariant
space is more than one, it becomes complicate to control a quantum
system from an arbitrary initial state to a given target state.
Nonetheless, by elaborately designing   the Lyapunov function, we
can still steer a quantum system to evolve into a desired state.

\section{Preparation  of edge state in AAH model}

The system of interest is the Fermi gas loaded in a 1D optical lattice, where the system Hamiltonian reads
\begin{eqnarray}\label{15}
H_{0}=-t\sum_{i=1}^{N-1}(c_{i}^{\dag}c_{i+1}+h.c.)+V\sum_{i=1}^{N}\cos(2\pi\alpha
i+\delta)c_{i}^{\dag}c_{i}.
\end{eqnarray}
Here $N$ is the total number of optical lattice sites. $t$ is the
hopping amplitude between the $i$-th and $(i+1)$-th site and
set to be the unit of energy throughout this paper.
$c_{i}$ and $c_i^{\dag}$ are the fermionic annihilation and creation
operators for the $i$-th site, respectively. $V$ denotes the strength of
commensurate potential with an rational number $\alpha$. When
$\alpha$ is an irrational number, $V$ represents the strength of
incommensurate potential which is the well-known
Aubry-Andr$\acute{\textrm{e}}$ model \cite{SA-AIPS3}, namely, all
eigenstates are extended (localized) for $V<2$ ($V>2$)  with a
single excitation.

\begin{figure}[h]
\centering
\includegraphics[scale=0.5]{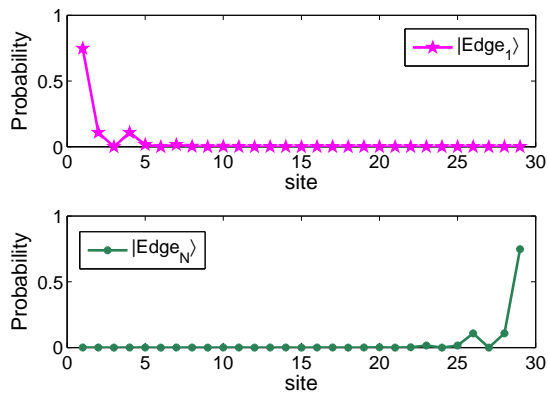}
\caption{(a) The probability amplitude of each site in edge state
$|\texttt{Edge}_1\rangle$. (b) The probability amplitude of each
site in edge state $|\texttt{Edge}_N\rangle$.
$|\texttt{Edge}_1\rangle$ ($|\texttt{Edge}_N\rangle$) denotes  the
edge state that the amplitude dominantly occupies at site $1$ ($N$).
We study in the single excitation subspace and other parameters
are chosen as $N$=29, $\alpha=1/3$, $V=1.5$, $\delta=2\pi/3$.} \label{fig:1}
\end{figure}

{Under  open boundary condition with proper parameters}, there exist edge
states $\{|\texttt{Edge}_i\rangle\}$ locating near boundary sites, which stems from the non-trivial
properties of the system and can be envisioned by mapping to the
2D quantum Hall effects in the Hofstadter problem
\cite{YEK1109,LJL-PRL108}. Especially, in the single excitation subspace, eigenstates localized near
boundaries can be found by resolving the eigenvalue equation,
\begin{eqnarray}\label{15}
H_{0}|\lambda_i\rangle=\lambda_i|\lambda_i\rangle, i=1,...,N,
\end{eqnarray}
where the $i$-th single particle eigenstate is given by $|\lambda_i\rangle=\sum_j^Nb_{j,i}c_j^{\dag}|0\rangle$.
$b_{j,i}$ are the superposition coefficients and satisfy
the normalization condition $\sum_{j}^{N}|b_{j,i}|^2=1$.
Figure \ref{fig:1} demonstrates the probability distribution of two edge states. Remarkably, the amplitude of edge states mainly locate at the boundary sites.

Next we explore how to use Lyapunov control to steer this system into one of two edge states. Note that one of the key points in the control process is how to choose a proper control Hamiltonian. If the structure of control Hamiltonian is much complex, even though it can be used to achieve edge states in theory, it may be difficult to manipulate in experiments. On the contrary, it may not steer the system into edge states when the control Hamiltonian is too simple. After combining with the feature of edge states, we find it is sufficient to prepare edge states by modulating the on-site energy of boundaries, which can be implemented easily. The control Hamiltonian then
can be written as
\begin{eqnarray}\label{15}
H_{1}=c_{1}^{\dag}c_{1},~~~H_{2}=c_{N}^{\dag}c_{N},
\end{eqnarray}
and the Lyapunov function can be chosen as
\begin{eqnarray}\label{15}
V_{1}=1-|\langle \psi_T|\psi(t)\rangle|^2,
\end{eqnarray}
which results in the control fields
$f_k(t)=A_{1k}\textrm{\textbf{Im}}[e^{i\arg\langle\psi(t)|\psi_T\rangle}
\langle \psi_T|H_k|\psi(t)\rangle].$  To be specific, we consider
$A_{1k}=1$ in numerical calculations, and
our goal is to steer an arbitrary given initial state to one of the
edge states, for instance, $|\psi_T\rangle=|\texttt{Edge}_N\rangle$.

\begin{figure}[htbp]
\centering
\includegraphics[scale=0.5]{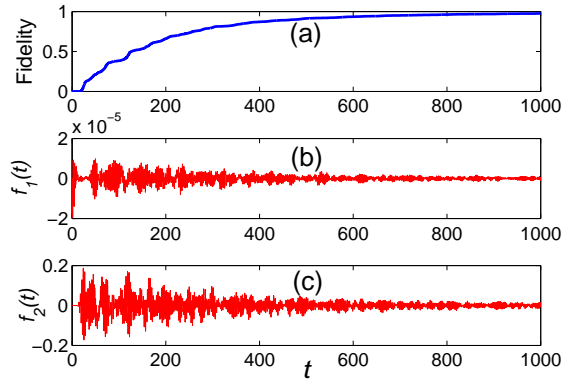}
\caption{ (a) The time evolution of fidelity
$F_{\psi(t)\texttt{Edge}_N}$. All parameters are in units of the
hopping amplitude $t$. (b) The time dependence  of the control field $f_1(t)$. (c)
The time dependence of the control field $f_2(t)$.} \label{fig:2}
\end{figure}

{One can find from} equation (\ref{2}) or (\ref{4}) that the
Lyapunov functions are time dependent via $|\psi(t)\rangle$, which
implies that the control fields would depend on the initial state at
the beginning. So, to design the control fields, we have to know the
initial state $|\psi_0\rangle$, though it might be  arbitrary. As
the single particle states (i.e., the single excitation occupies
only at site $n$, denoted by $|n\rangle (n=1,...,N)$) are more easily
to prepare, we consider these states as initial states in the
following. Note that this method can also be applied for
arbitrary superposition of the single particle states. Figure
\ref{fig:2} shows the time evolution of the  fidelity defined by
$F_{\psi(t)\texttt{Edge}_N}=|\langle\psi(t)|\texttt{Edge}_N\rangle|$
and the control fields $f_k(t)$ while the initial state
of system is chosen as $|\psi_0\rangle=|3\rangle$. We find that the
control fields $f_k(t)$ can steer the initial state to the edge state
eventually. In addition to this, it can also be observed that the
control field $f_2(t)$ plays an important role in the evolution,
since the control field $f_1(t)$ is comparatively  small. The
physics behinds this result can be understood  as follow. The
excitation mainly occupies at site $N$ for the target state
$|\texttt{Edge}_N\rangle$, thus the control Hamiltonian $H_2$
dominates during the time evolution. As a result, the corresponding
control field plays an important role.

\begin{figure}[b]
\centering
\includegraphics[scale=0.5]{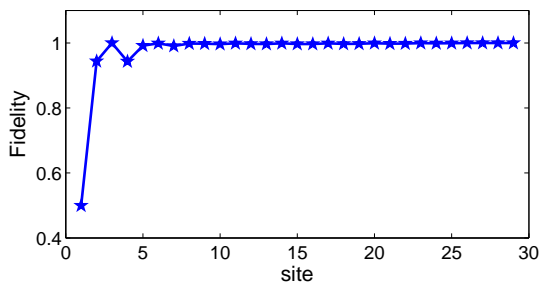}
\caption{The fidelity $F_{\psi_s\texttt{Edge}_N}$ versus different initial
states, $|\psi_0\rangle=|n\rangle~(n=1,...N)$. The other parameters
are the same as in figure \ref{fig:2}.} \label{fig:3}
\end{figure}

\begin{figure}[htbp]
\centering
\includegraphics[scale=0.45]{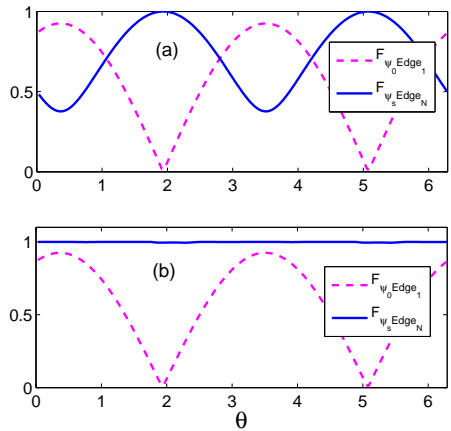}
\caption{$F_{\psi_s\texttt{Edge}_N}$ and $F_{\psi_0\texttt{Edge}_1}$
as a function of $\theta\in[0, 2\pi]$  with (a) Lyapunov function
$V_1$ and (b) Lyapunov function $V_2$. The initial state is chosen
as $|\psi_0\rangle=\cos{\theta}|1\rangle+\sin{\theta}|2\rangle$
because the amplitude of the edge state $|\texttt{Edge}_1\rangle$
mainly locates at site 1 and site 2.} \label{fig:4}
\end{figure}

Next, we investigate the fidelity $F_{\psi_s\texttt{Edge}_N}$
(denotes the fidelity between the final state $|\psi_s\rangle$ and
the target state $|\texttt{Edge}_N\rangle$) at different initial
states $|\psi_0\rangle=|n\rangle (n=1,...,N)$, which are illustrated
in figure  \ref{fig:3}. There is an
intuitive hypothesis from figure \ref{fig:3} that the fidelity $F_{\psi_s\texttt{Edge}_N}$
might be related to the fidelity $F_{\psi_0\texttt{Edge}_1}$
(denotes the fidelity between the initial state $|\psi_0\rangle$ and
the edge state $|\texttt{Edge}_1\rangle$). In order to confirm it,
figure \ref{fig:4}(a) demonstrates the relation between the fidelity
$F_{\psi_s\texttt{Edge}_N}$ and the fidelity
$F_{\psi_0\texttt{Edge}_1}$. We can see in figure \ref{fig:4}(a)
that steering the system to the target state becomes much difficult
with the  increasing of $F_{\psi_0\texttt{Edge}_1}$, since the edge
state $|\texttt{Edge}_1\rangle$ is also an element of the LaSalle's
invariant set (we will demonstrate it in the following).  Therefore,
small value of $F_{\psi_0\texttt{Edge}_1}$ benefits the control
process when we set the edge state $|\texttt{Edge}_N\rangle$ as  the
target state. This observation also holds true when the initial
state is an arbitrary superposition of the single particle
states, which is not shown in figure \ref{fig:4}(a). Those results
can be explained as follow. As the initial state can be rewritten as
a superposition of the eigenstates of the system Hamiltonian $H_0$, i.e.,
\begin{eqnarray}\label{15}
|\psi_0\rangle=a(0)|\texttt{Edge}_1\rangle+\sum_{i=2}^{N}a_i(0)|\lambda_i\rangle.
\end{eqnarray}
Since the edge state $|\texttt{Edge}_1\rangle$ is  an element of the
largest invariant set, the amplitude of the edge state
$|\texttt{Edge}_1\rangle$ remains almost unchanged during the
control process and the other eigenstates would be steered into the
target state. Consequently, it can not be steered into the target
state perfectly if the initial state contains the component of the
edge state $|\texttt{Edge}_1\rangle$. The details of proof can be
found in the appendix.

\begin{figure}[b]
\centering
\includegraphics[scale=0.5]{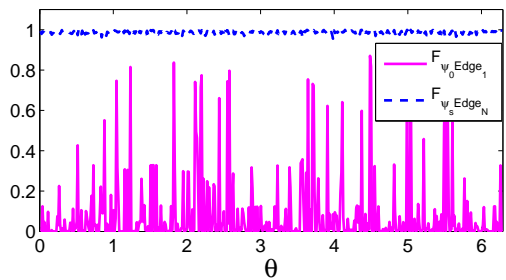}
\caption{$F_{\psi_s\texttt{Edge}_N}$ (blue line) and
$F_{\psi_0\texttt{Edge}_1}$ (pink line) as a function of  parameter
$\theta$ with   Lyapunov function $V_2$ and  operator $P_1$. The
initial state is
$|\psi_0\rangle=\cos{\theta}|n\rangle+\sin{\theta}|m\rangle$, where
$n, m$ are stochastic integer created between   1 and  $N$, while
$\theta$ $\in[0, 2\pi]$, $N=29$. The results are a collection of 500
random initial states.} \label{fig:7}
\end{figure}

As the choice of Lyapunov function is not unique, we  can choose
another Lyapunov function to design the control fields for the
system, which we have elucidated in Sec. II.
Based on several trials, we choose the operator $P$ as
\begin{eqnarray}\label{15}
P_1=p_f|\texttt{Edge}_N\rangle\langle\texttt{Edge}_N|+\sum_{i\neq
f}p_i|\lambda_i\rangle\langle\lambda_i|,
\end{eqnarray}
where $p_i=\lambda_i$ and $p_f=-3$ guarantee
$p_f<p_i$. The corresponding control fields are specified as
$f_k(t)=-A_{2k}^1\langle \psi(t)|i[H_k,P_1]|\psi(t)\rangle$ with
$A_{2k}^1=5$. From figure \ref{fig:4}(b), we observe that a high
fidelity of final state can be obtained with an (almost)
arbitrarily given initial state by elaborately
designing the value of $p_i$ ($i=1,...,N$) in equation (\ref{Lya2}) \cite{note}. More generally, figure
\ref{fig:7} shows  a collection of  fidelity with an arbitrary given
initial state in the form
$|\psi_0\rangle=\cos{\theta}|n\rangle+\sin{\theta}|m\rangle$, and we
find that it can approach 98.74\% on average. Here $n$ and $m$ in
the initial states are integers  stochastically created from 1
to $N$. This observation argues that the value of
$F_{\psi_0\texttt{Edge}_1}$ has slightly effects on the fidelity of
final state when  the operator $P$ is elaborately constructed.

The above numerical calculations imply  that different  Lyapunov
functions lead to different fidelity and behaviors of convergence
due to the distinct largest invariant set. To be more specific, The
largest invariant set with Lyapunov function $V_1$ is $\mathcal
{E}_1=\{|\psi\rangle, \
 \ |\langle\psi_T|H_k|\psi\rangle=0, k=1,2\}$ while
it is $\mathcal {E}_2=\{|\psi\rangle,\ \ |\langle\lambda_{i}
|H_k|\lambda_j\rangle\langle\lambda_j|\psi\rangle\langle\psi|
\lambda_i\rangle=0, k=1,2; ~i,j=1,...,N\}$ for the Lyapunov function
$V_2$ \cite{SK-A44}. It can be verified easily that only the state
$|\texttt{Edge}_N\rangle$ belongs to the largest invariant set when
designing by the Lyapunov function $V_2$ and the hermitian operator
$P_1$. Nevertheless, this is not the case for the Lyapunov function
$V_1$. There exists another solution
$|\psi\rangle=|\texttt{Edge}_1\rangle$ that satisfies the condition
$\langle\psi_T|H_k|\psi\rangle=0(k=1,2)$, thus the  largest
invariant set $\mathcal {E}_1$ is spanned by
$\{|\texttt{Edge}_1\rangle,|\texttt{Edge}_N\rangle\}$. This enlarged
invariant set caused by the Lyapunov function $V_1$ makes the
control behaviors different from that of Lyapunov function $V_2$. As
a consequence, the control with the Lyapunov function $V_1$ may not
steer the system to the target state very well. From this point of
view, it is important to choose a proper Lyapunov function in order
to obtain a high fidelity of the target state.

Since there are many ways to choose Lyapunov functions as aforementioned,
it gives rise to a question: how to optimize the
Lyapunov function in order to obtain a high fidelity of final state.
In the following, we analyze and answer this question. The
Lyapunov function is required to be real and positive, so it can be
written as an average of hermitian operator $P$. Therefore, the
Lyapunov function in equation (\ref{4}) is general form. By the principle of Lyapunov
control, control fields must vanish and Lyapunov
function should reach its minimum when the system arrives at  the
target state. This  requires that the operator $P$ must commute with
the system Hamiltonian $H_0$ (see equation (\ref{5})). Hence the
general form of operator $P$ takes,
\begin{eqnarray}\label{15}
P=p_0\cdot\mathcal {I}+\sum_i^{N}
p_i|\lambda_i\rangle\langle\lambda_i|,
\end{eqnarray}
where $|\lambda_i\rangle$ ($i=1,...,N$) are the eigenstates of system
Hamiltonian and $p_i(i=0,...,N)$ are arbitrary real numbers.

\begin{figure}[b]
\centering
\includegraphics[scale=0.5]{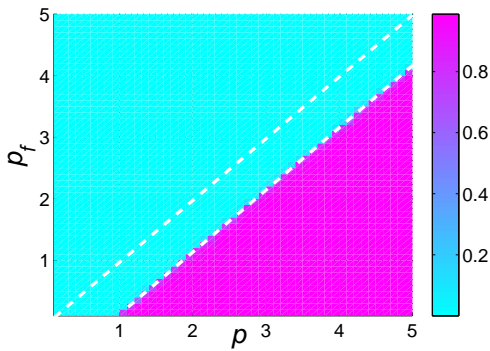}
\caption{ The fidelity of final state to the edge state
$|\texttt{Edge}_N\rangle$ as a function of coefficients $p$ and
$p_f$. We have chosen the initial state $|\psi_0\rangle=|3\rangle$,
the control time $t=1000$, and the Lyapunov function $V_2$ with the
operator $P_1$.} \label{fig:01}
\end{figure}

\begin{figure}[htbp]
\centering
\includegraphics[scale=0.5]{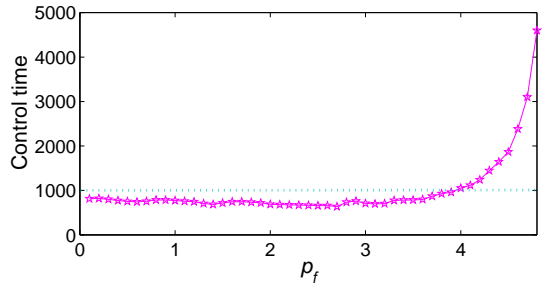}
\caption{ The control time as a function of $p_f$ (pink line with
stars) while the fidelity reaches 0.97 and we have fixed $p=5$. The
other parameters are the same as in figure \ref{fig:01}.}
\label{fig:011}
\end{figure}

With these considerations, the optimalization over the Lyapunov
function reduces to searching a set of $\{ p_i \}$ that maximize the
fidelity of final state. Note that the term with identity operator
$\mathcal {I}$ is  a constant when substituting $P$ into equation
(\ref{4}). It does not affect the designing for control fields, as a
consequence, $p_0$ can be an arbitrary value. To simplify the
optimalization scheme, we assume  $p_i=p$ $(i\neq f)$, and the
optimalization is furthermore simplified to   analyze the
relationship between the fidelity and the coefficients $p$ and $p_f$
with a fixed control time. It should be addressed that the
optimalization should be taken over all $p_i$ ($i=1,...,N$), but it
is a consuming task when $N$ is large, so here we simplify the
problem and use the simple case to exemplify the optimalization
scheme. Figure \ref{fig:01} shows the fidelity of final state as a
function of $p_f$ and $p$. We find that it fails to prepare the edge
state when $p_f>p$ since the final state is not the target state
when the Lyapunov function $V_2$ reaches its minimum. We can also
see that the fidelity of final state is almost unchanged with
$p-p_f>1$ and vanishes with $p-p_f<1$. The reason can be found in
figure \ref{fig:011}, which shows the relation between the control time
and $p_f$, where the control process is completed when the fidelity
of final state reaches 0.97 with $p=5$. For $p-p_f<1$, it takes a
long time to have such a fidelity. In other words, the fidelity of
final state is very small if the control time stops  at $t=1000$. In
addition, we find from figure \ref{fig:011} that the control time
almost stays at a fixed value for  $p-p_f>1$, which provides us with
a guidance for designing the coefficients in the Lyapunov function.

\section{robustness}

\begin{figure}[h]
\centering
\includegraphics[scale=0.5]{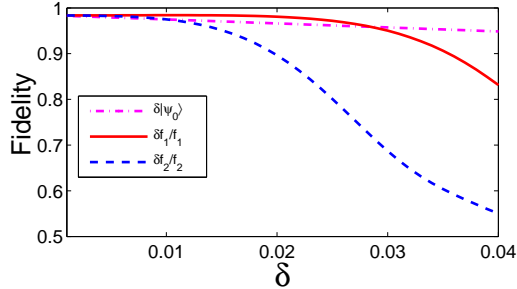}
\caption{ We have assumed that the perfect initial state is
$|3\rangle$, where the errors caused by unperfect preparation is
defined by $\delta=1-|\langle\psi_0|3\rangle|$. The red solid line and
blue dash line shows the fidelity of final state versus the errors
caused by mismatching control fields $f_1(t)$ and $f_2(t)$
respectively. The other parameters are the same as in figure
\ref{fig:01}. } \label{fig:02}
\end{figure}

In practise, there inevitably exist some errors during the control process.
For example, it requires to know the
initial state precisely, but the preparation of initial state
may not be perfect. Errors may also occur  in the
application of control fields. In the following, we investigate the
effect of these errors on the fidelity of final state, which are
characterized by $\delta$. Namely, if the theoretical exact values of
control fields are $f_k(t)$, the actual control fields applied
in the control process are $f^{\prime}_k(t)=(1+\delta)f_k(t)$. As to
errors in the initial state, we randomly create  a state, and mix it
with the initial state (e.g., $|3\rangle$) such that
 $\delta=1-|\langle\psi_0|3\rangle|$, where $|\psi_0\rangle$ is the
actual initial state and $|3\rangle$ is the ideal initial state. With
these considerations, we simulate the influence of errors on the fidelity of final state
in figure \ref{fig:02}. It can be found that the unperfect
preparation of initial state $|3\rangle$ does not have a serious
effect on the fidelity of final state. The control field $f_1(t)$
has a  slight influence on the fidelity of final state, but the
control field $f_2(t)$ does affect the fidelity of final state. The
reason is that the edge state $|\texttt{Edge}_N\rangle$ is mainly
localized at the site $N$ in the optical lattice, then the control
field $f_2(t)$ dominates in the control process.

\begin{figure}[h]
\centering
\includegraphics[scale=0.33]{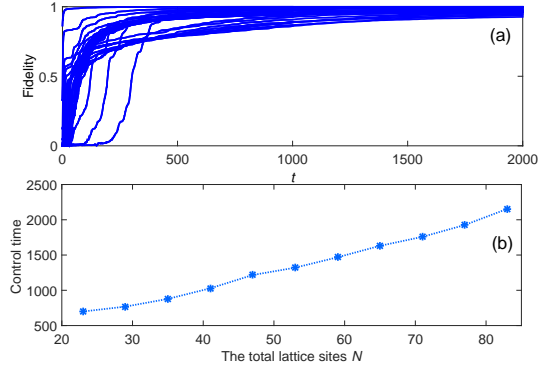}
\caption{(a) The dynamics evolution of fidelity with different
initial states $|\psi_0\rangle=|n\rangle,~~n=1,...,N$. The Lyapunov
function is $V_2$ with the operator $P_1$. (b) The control time
versus different total lattice sites $N$ when the fidelity is
0.99, where the initial state is $|3\rangle$.} \label{fig:03}
\end{figure}

On the other hand, it can be found in figure \ref{fig:03}(a) that different initial
states manifest distinct dynamics behaviors and convergence time in
the control process since the Lyapunov functions rely on the initial
state $|\psi_0\rangle$ through $|\psi(t)\rangle$. In addition, the
total lattice sites can also affect the control time and the
time for edge state preparation is approximate linearly proportional
to the total lattice sites, as shown in figure
\ref{fig:03}(b).

\section{Boundary-boundary entangled states}

For the system under Lyapunov control, we have shown that the
LaSalle's invariant set is spanned by $\{|\texttt{Edge}_1\rangle,
|\texttt{Edge}_N\rangle\}$ for Lyapunov function $V_1$. As a result
the state of the system would asymptotically converge to
\begin{eqnarray}\label{16}
|\psi(t)\rangle=\alpha|\texttt{Edge}_1\rangle+\beta|\texttt{Edge}_N\rangle.
\end{eqnarray}
Equivalently, it can be rewritten as
\begin{eqnarray}\label{15}
 |\psi(t)\rangle=\sum_{n=1}^{N}C_n(t)|n\rangle,
\end{eqnarray}
where $C_n(t)$ represents the probability amplitude of a single
particle located at the $n$-th site in the optical lattice. In order to
get a high degree of boundary-boundary entangled states, we
deform the Lyapunov function as
\begin{eqnarray}\label{15}
V_3=1-|\langle
\texttt{Edge}_1|\psi(t)\rangle|^2-|\langle\texttt{Edge}_N|\psi(t)\rangle|^2,
\end{eqnarray}
and the time derivative of $V_3$ yields
\begin{eqnarray}\label{15}
\dot{V}_3=-2\sum_{k=1,2}f_k(t)\sum_{m=1,N}|\langle
\psi(t)|\texttt{Edge}_m\rangle|\textrm{\textbf{Im}}[e^{i\arg{\langle\psi(t)|\texttt{Edge}_m\rangle}}
\langle \texttt{Edge}_m|H_k|\psi(t)\rangle].
\end{eqnarray}
Hence, the control field  can be  chosen as
$f_k(t)=-A_{3k}\sum_{m=1,N}\textrm{\textbf{Im}}[e^{i\arg\langle
\psi(t)|\texttt{Edge}_m\rangle} \langle
\texttt{Edge}_m|H_k|\psi(t)\rangle]$ with $A_{3k}=1$.

When investigating the entanglement between site $1$ and site
$N$ in the optical lattice, the remaining sites should be traced
out, i.e.,
$\rho_{1N}(t)=\textrm{Tr}_{2...N-1}(|\psi(t)\rangle\langle\psi(t)|)$.
In the Hilbert space spanned by $\{|00\rangle, |10\rangle,
|01\rangle, |11\rangle\}$, we have the following expression for
reduced density matrix $\rho_{1N}(t)$
\begin{equation}
{\rho _{1N}(t)}=\left( \begin{array}{cccc}
   a & 0 & 0 & 0  \\
   0 & b & d & 0  \\
   0 & d^* & c & 0  \\
   0 & 0 & 0 & 0  \\
\end{array} \right)
\end{equation}
with $a=\sum_{n=2}^{n=N-1}|C_n|^2$, $b=|C_1|^2$, $c=|C_N|^2$,
$d=C_1C_N^*$. Then we use concurrence \cite{WKW-PRL80} to
measure the degree of entanglement,
\begin{eqnarray}\label{15}
\mathcal {C}(\rho)\equiv \max\{0,
\lambda_1-\lambda_2-\lambda_3-\lambda_4\},
\end{eqnarray}
where $\lambda_l~(l=1,2,3,4)$ are the square roots of eigenvalues of
the   matrix
$\rho(\sigma_y\otimes\sigma_y)\rho^{\ast}(\sigma_y\otimes\sigma_y)$
in decreasing order, and $\rho^{\ast}$ is the complex conjugate of
$\rho$. According to the expression of ${\rho _{1N}}(t)$, it is
straightforward to find that
\begin{eqnarray}\label{15}
\mathcal {C}_{1N}=\max\{0, 2\sqrt{bc}, 2|d|\}.
\end{eqnarray}

\begin{figure}[h]
\centering
\includegraphics[scale=0.5]{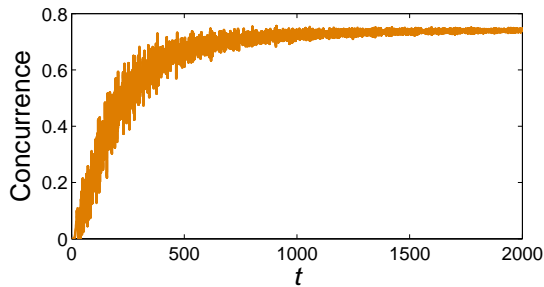}
\caption{The  concurrence
$\mathcal {C}_{1N}$ as a function of time $t$, where Lyapunov
function is $V_3$.}
\label{fig:8}
\end{figure}

\begin{figure}[htbp]
\centering
\includegraphics[scale=0.5]{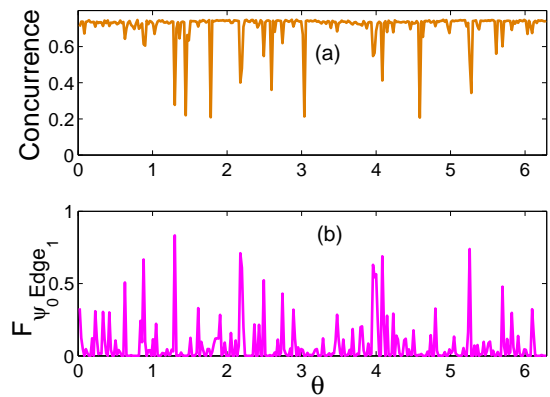}
\caption{(a) The concurrence $\mathcal {C}_{1N}$ and (b) the fidelity
$F_{\psi_0\texttt{Edge}_1}$ as a function
of initial states characterized by the parameter $\theta$, where
Lyapunov function is $V_3$, $\theta\in[0, 2\pi]$. The initial state is
$|\psi_0\rangle=\cos{\theta}|n\rangle+\sin{\theta}|m\rangle$, where
$n, m$ are stochastic integer ranging from 1 to $N$. We simulate
the concurrence and fidelity with 300 random initial states.}
\label{fig:9}
\end{figure}

Figure \ref{fig:8} illustrates the dynamics behavior of concurrence
with initial state $|\psi_0\rangle=|3\rangle$, which shows that
the  concurrence $\mathcal {C}_{1N}$ approaches to a fixed value
0.74. It should be noted that the entanglement of steady state
depends on the initial state, in particular it is very
closely connected to the fidelity $F_{\psi_0\texttt{Edge}_1}$
between the initial state and the edge state
$|\texttt{Edge}_1\rangle$, which is shown in figure \ref{fig:9}.
Additionally, it is readily found that the average value of concurrence
reaches about 0.71 in numerical calculations. The reason why the concurrence of
steady entangled state is below 0.75 can be found
as follow. As we have shown, the final state is approximate a
superposition of two edge states and the probability amplitudes
of edge states on the first and end site are about 0.7474 in figure
\ref{fig:1}. When $\alpha=\beta=\frac{1}{\sqrt{2}}$ in equation
(\ref{16}), the entanglement of edge states  reaches its maximum,
which helps  to establish the concurrence of boundary-boundary
entangled state. Although the boundary-boundary entanglement is not
very high, it is symmetry protected and therefore maybe useful in
quantum information processing (actually, it is almost a maximal
entanglement between two edge states).

\section{conclusion}

In conclusion, we have proposed a scheme to prepare edge states in
the optical lattice by shaking boundaries, which is motivated
by the fact that the edge states exhibit interesting physical
properties. The control Hamiltonian we use is restricted to the boundaries of the
optical lattice and the control field is designed by Lyapunov control where the
Lyapunov function lies at the heart of control design. By
choosing different Lyapunov functions, we have shown that the control process
would lead to different fidelity and convergence feature.
It have been found that the edge state we obtain depends on
the initial states with Lyapunov function $V_1$ while it can be
obtained for almost arbitrary initial states when using Lyapunov
function $V_2$ with operator $P_1$. In addition, we have discussed
the influence of errors on the fidelity of edge state.
By this proposal, we can also prepare boundary-boundary
entangled state which is actually the maximum superposition of
two edge states.

\ \ \\
\ \ This work is supported by the National Natural Science Foundation of
China (Grants No. 11534002, No. 61475033 and 11475037), and supported by
the Fundamental Research Funds for the Central Universities under
grant No. DUT15LK26.

\section*{APPENDIX}

In this appendix, we show that the amplitude of edge state
$|\texttt{Egde}_1\rangle$ almost remains unchanged under the
control   with Lyapunov function $V_1$.  In the following, we adopt
the same notions as those in the main text.

Suppose the initial state of this system is
\begin{eqnarray}\label{15}
|\psi_0\rangle&=&a(0)|\texttt{Edge}_1\rangle
+\sum_{i=2}^{N}a_i(0)|\lambda_i\rangle.
\end{eqnarray}
The state at time $t$ becomes
\begin{eqnarray}\label{15}
|\psi(t)\rangle=a(t)|\texttt{Edge}_1\rangle+\sum_{i=2}^{N}a_i(t)|\lambda_i\rangle.
\end{eqnarray}
According to the Schr\"odinger equation, we can deduce the differential equation
for the probability amplitude of edge state
$|\texttt{Egde}_1\rangle$ ($\hbar=1$),
\begin{eqnarray}\label{15}
i\cdot\dot{a}(t)&=&\langle\texttt{Edge}_1|[H_0+\sum_{k=1}^2f_k(t)H_k]|\psi(t)\rangle   \nonumber\\
&=&\lambda_{1}\cdot a(t)+\gamma_{1}\cdot a(t)+\sum_{j=2}^{N}
\gamma_j\cdot a_j(t),
\end{eqnarray}
where
$\gamma_j=\langle\texttt{Edge}_1|[f_1(t)H_1+f_2(t)H_2]|\lambda_{j}\rangle,~j=2,...,N.$

On the other hand, since the edge state
$|\texttt{Egde}_N\rangle$ locates near at site $N$, the probability amplitude of edge state
$|\texttt{Egde}_N\rangle$ on the site $1$ almost vanishes, i.e., $\epsilon\ll1$
(the magnitude order $\epsilon\sim10^{-5}$ in the case of $N=29$).
It is similar to the situation of edge state $|\texttt{Egde}_1\rangle$.
Thus one can estimate the order of magnitude in
$H_1|\texttt{Egde}_N\rangle$ and $H_2|\texttt{Egde}_1\rangle$:
\begin{eqnarray}\label{15}
H_1|\texttt{Egde}_N\rangle&=&H_1|\psi_T\rangle\sim\epsilon\cdot(1,\underbrace{0,...,0}_{N-1})^T,  \nonumber\\
H_2|\texttt{Egde}_1\rangle&\sim&\epsilon\cdot(\underbrace{0,...,0}_{N-1},1)^T,
\end{eqnarray}
where the superscript $T$ represents transposition and the order of
magnitude in the function $f_1(t)$ can be estimated as well:
\begin{eqnarray}\label{15}
f_1(t)=\texttt{\textbf{Im}}[e^{i\textrm{arg}\langle\psi(t)|\psi_T\rangle}\langle\psi_T|H_1|\psi(t)\rangle]
\sim \epsilon.
\end{eqnarray}
Those estimations give the order of magnitude in $\gamma_j$:
\begin{eqnarray}\label{15}
|\gamma_j|&=&|f_1(t)\langle\texttt{Egde}_1|H_1|\lambda_{j} \rangle+f_2(t)\langle\texttt{Egde}_1|H_2|\lambda_{j}\rangle|
 \sim\epsilon\ll|\lambda_{1}|.
\end{eqnarray}
Therefore the differential equation of $a(t)$ can be approximated as
\begin{eqnarray}\label{15}
i\cdot\dot{a}(t)\simeq\lambda_{1}\cdot a(t),
\end{eqnarray}
leading to
\begin{eqnarray}\label{15}
|a(t)|^2\simeq constant=|a(0)|^2.
\end{eqnarray}
Hence, the amplitude of edge state $|\texttt{Egde}_1\rangle$ almost remains
unchanged  during the control process with Lyapunov
function $V_1$.


\begin{references}
\bibitem{note2}We   get   $V_1$ in equation (\ref{2})
by taking $P=\mathcal {I}-|\psi_T\rangle\langle \psi_T|$ in equation
(\ref{4}), where $\mathcal {I}$ is the identity operator.
\bibitem{note}The Lyapunov control can not steer a system from an
initial state in the LaSalle's invariant set to the target state.
\end{references}
\end{document}